\documentstyle[psfig]{aa}
\def\I{\'{\i}}
\def\h2{\hbox{H$_2$}}
\def\c18o{\hbox{C$^{18}$O}}

\def\m13co{\hbox{~$^{13}$CO}}
\def\NH3{NH$_3$}
\def\kms{~km~s$^{-1}$~}
\def\cmmt{~cm$^{-3}$~}
\def\cmmd{~cm$^{-2}$~}

\def\mtres{M+3.06+0.34~}

\def\muno{M+1.56$-$0.30~}
\def\le{$\leq$}

\def\le{$\leq$}

\def\nh2{\hbox{$n_{\rm H_2}$}}
\def\Nh2{\hbox{$N_{\rm H_2}$}}   
\newcommand{\gsim}{\raisebox{-.4ex}{$\stackrel{>}{\scriptstyle \sim}$}}
\newcommand{\lsim}{\raisebox{-.4ex}{$\stackrel{<}{\scriptstyle \sim}$}}

\begin{document}
\thesaurus{08(10.03.1; 13.09.3; 13.09.4; 09.13.2; 09.03.1; 09.04.1)}
\title{Non-equilibrium H$_2$ ortho-to-para ratio in two
molecular clouds of the Galactic Center\thanks{ISO is an ESA project with instruments funded by ESA
Member States (especially the PI countries: France, Germany,
the Netherlands and the United Kingdom) and with the
 participation of ISAS and NASA.}}
\author{N.J.~Rodr\'{\i}guez-Fern\'andez\inst{1}, J.~Mart\'{\i}n-Pintado\inst{1},
P.~de Vicente\inst{1}, A.~Fuente\inst{1},  
S.~H\"uttemeister\inst{2},  T.L.~Wilson\inst{3,4}, \and D.~Kunze\inst{5}}

\institute{Observatorio Astron\'omico Nacional (IGN), Apartado 1143,
E-28800, Alcal\'a de Henares, Spain (nemesio, martin, vicente, fuente$@$oan.es)
\and Astronomische Institute, Auf dem Huegel 71, D-53121 Bonn 1, Germany
(huette@astro.uni-bonn.de)
\and Max--Planck Institut f\"ur Radioastronomie, Postfach 2024, D 53010 Bonn,
\and Sub-mm Telescope Observatory, Steward Observatory, The University
of Arizona, Tucson, Az, 85728, USA \\ (twilson@as.arizona.edu)
\and MPE, Giesenbachstrasse 1, D-85748 Garching, Germany (kunze@mpe.mpg.de)}

\date{Received date; accepted date}

\maketitle

\markboth{ N.J. Rodr\'{\i}guez-Fern\'andez et al.: Non-LTE OTPR in two 
GC clouds}{ }

\begin{abstract}
We present ISO observations of the S(0), S(1), S(2), and S(3) rotational lines
of molecular hydrogen from  two molecular clouds near              
the Galactic Center (GC). We have also measured continuum dust
emission at infrared wavelengths with ISO and  
the rotational radio lines J=1--0 of $^{13}$CO and C$^{18}$O
 and J=2--1 of  C$^{18}$O 
with the IRAM-30m telescope.  Using the dust continuum spectra
and the CO lines we derive a total visual extinction of $\sim$ 15-20 magnitudes 
toward these GC clouds.     
After correcting the H$_2$ data for extinction, the gas 
temperatures are $\sim$ 250~K
and the column densities of warm gas are $\sim 2\times 10^{21}$~cm$^{-2}$. 
This is the first direct measure of the H$_2$ column
 densities of the warm component;
with this, we  estimate an NH$_3$ abundance in the warm gas   
 of $\sim 2~10^{-7}$. 
The column density of warm gas is, at least, a factor of 100 larger than
the corresponding column densities derived from the  warm dust.
The observed ortho-to-para ratio (OTPR) is  $\sim$ 1, clearly below the
local thermodynamical equilibrium (LTE) OTPR for gas at 250 K of $\sim 3$.
Low velocity shocks ($\sim$ 10 km s$^{-1}$) 
are the most likely explanation for the column densities of 
warm gas and dust and the non-LTE H$_2$ OTPR. 

\end{abstract}


\section{Introduction}

The central $\sim 6^{\circ}$ of our galaxy exhibit a large 
accumulation of molecular material
which is forming big molecular clouds whose 
masses and sizes are so large as
10$^6$ M$_{\odot}$ and 15~pc, respectively. These clouds 
are denser (average
densities of 10$^4$ \cmmt), more turbulent 
(line widths of $\sim$ 20 \kms), and hotter
(with a warm component with temperatures, $T$, up to 200-300 K) 
than the clouds of 
the disk of the galaxy (see e.g. Morris \& Serabyn \cite{morris}).
The high temperatures in Galactic Center (GC) clouds were known 
basically by observations of \NH3 inversion 
lines over limited regions (G\"usten et al. \cite{gusten81};
Mauersberger et al. \cite{mauersberger}).
H\"uttemeister et al. (\cite{huttemeister93}) analyzed  36
molecular clouds distributed all along the Central Molecular Zone 
and the ``Clump 2" complex;
they showed that high kinetic temperatures  are a general characteristic of
the GC clouds and not only of those located
close to Sgr A and Sgr B2. 
In the disk of the galaxy, kinetic temperatures higher than 100 K are 
associated with infrared sources, that is, embedded stars which heat the dust 
and subsequently the gas by collisions with the dust grains.
The typical
sizes of such  regions are less than 1 pc. The high  kinetic
temperatures in the GC clouds are found in      regions of $\sim$ 10 pc,
where one measures large column densities of cold dust
 ($T <$ 30 K, Odenwald \& Fazio \cite{odenwald};
Cox \& Laureijs \cite{cox}). This rules out gas-dust collisions as a possible
heating mechanism of the warm component. Dissipation of turbulence due to
shocks induced by the rotation of the galaxy  could be the main
heating mechanism in the GC clouds (Wilson et al. \cite{wilson82}).

Unfortunately the NH$_3$ abundance  in the warm
component  was  unknown since  one could not estimate the warm
\h2 column densities.
The {\em Infrared Space Observatory} (ISO; Kessler et al. \cite{kessler}),
has allowed  us, for the first time, to measure 
directly the total column density
of warm gas by observing pure-rotational  lines of \h2. These
trace gas with temperatures of a few hundreds Kelvin.
  ISO has also allowed us to study the \h2 ortho-to-para ratio (OTPR), 
which can help determine       the possible heating mechanism
and the origin of this molecule.
Before  ISO, the \h2 OTPR
had been studied in regions with temperatures of $\sim$ 2000~K, using
the vibrational lines. In such shock-excited sources, one measures         
an OTPR of $\sim$ 3 (Smith et al. \cite{smith}),
which is the local thermodynamical equilibrium (LTE) value for $T \gsim$ 200 K.
In contrast, for    regions heated mainly by ultraviolet (UV) radiation
(Photodissociation regions [PDRs]), the vibrational lines give     OTPRs in
the range of 1.2-2 (see e.g. Chrysostomou et al. \cite{chrysostomou}).
However, these low OTPRs  might  not be
a consequence of an actual non-LTE ortho-to-para {\em abundances} ratio but
a result of optical depth effects in the fluorescence-pumping of the
ortho-\h2 (Sternberg \& Neufeld \cite{sternberg}). Using these considerations,
one can  explain
why the PDR in S140 exhibits an OTPR $\sim$ 2 in the vibrational states
but  3 in the lowest rotational levels.

There are two cases of non-equilibrium OTPR measured 
from the pure-rotational lines: the shock excited source HH54
(Neufeld et al. \cite{neufeld}) and the PDR 
associated with the reflection nebula
NGC 7023 (Fuente et al. \cite{fuente99}). The first case has been
explained using     the shocks model of Timmermann (\cite{timmermann}),
which involves transient heating by low velocity shocks.
To explain  the non-LTE OTPR in NGC 7023, it was necessary to
invoke a dynamic dissociation front.

To investigate the thermal balance of the GC clouds
we have selected 18 clouds from the samples of H\"uttemeister et al. 
(\cite{huttemeister93}) and Mart\I n-Pintado et al. (\cite{mp97})
and we have observed them with the ISO satellite. 
In this paper we present \h2 observations toward two 
sources which show similar characteristics 
(also shown in the NH$_3$ studies of                         
H\"uttemeister et al. \cite{huttemeister93}), indicating that 
their heating mechanisms are also very similar.
In particular, they show a non-LTE OTPR.
The detection of OTPRs out of equilibrium 
in the GC clouds gives us new insights into
the heating mechanism, since the gas must be
heated to  several hundreds K almost without changing
the OTPR of cold gas.

In Sects. 2 and 3,  we present observations and results, respectively, and in 
Sect. 4 we discuss the possible heating mechanism and the 
origin of the non-equilibrium OTPR.

\begin{figure*}[p]
\centerline{\psfig{figure=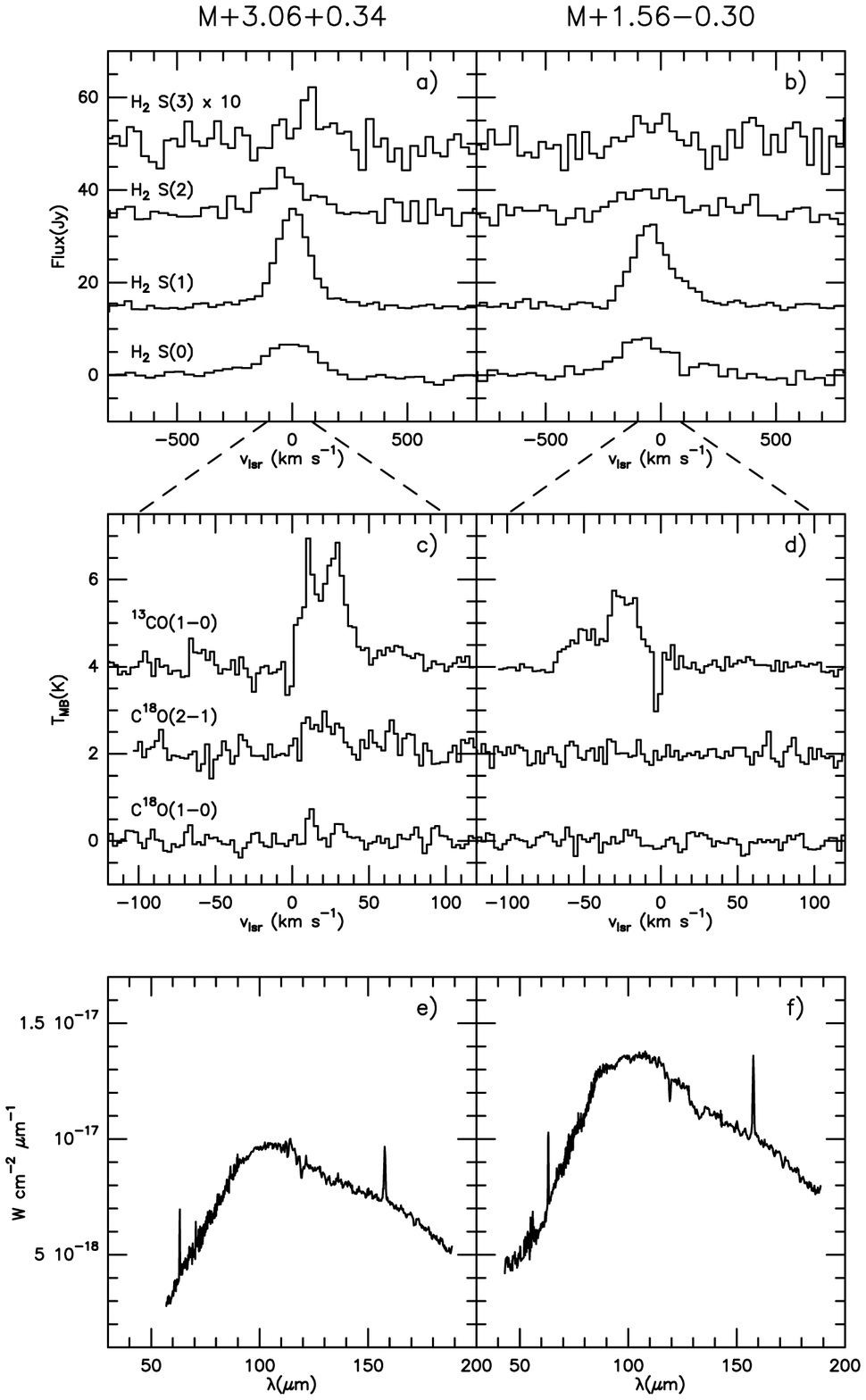,bbllx=69pt,bblly=130pt,bburx=490pt,bbury=753pt,height=18cm}}
\caption[]{ Spectra of the two sources: 
  {\bf a-b} \h2  spectra   taken with the SWS. 
  {\bf c-d} IRAM-30m spectra of the \c18o(1--0), \m13co(1--0),
       and \c18o(2--1) lines.
  {\bf e-f} LWS full grating spectra.      
            Note the different radial velocities ranges in Fig. 1a-b
            and Fig. 1c-d.
   }
\label{figobs}
\end{figure*}         
\section{Observations and data reduction}
\subsection{ISO observations}
 We  observed  the  pure-rotational  H$_2$ lines
 S(0), S(1), S(2), and S(3) with the {\em Short Wavelength 
Spectrometer} (SWS;  de Graauw et al. \cite{graauw})  on board  ISO
toward two      molecular clouds.
The galactic coordinates and ISO beam sizes are given in Table 1. 
These sources are among the farthest from the dynamical center
of the galaxy in our sample.
\mtres [$\alpha$(2000)=  17$^{\rm h}$ 51$^{\rm m}$ 26\fs 4,
$\delta$(2000) = -26\degr 08\arcmin 29\farcs 4]
is located in the  ``Clump 2" complex (Stark \& Bania \cite{stark}), 
while     M+1.56$-$0.30 
[$\alpha$(2000)=  17$^{\rm h}$ 50$^{\rm m}$ 26\fs 6, 
$\delta$(2000) = -27\degr 45\arcmin 29\farcs 5]
belongs to the ``$l$=1\fdg5-complex" (Bally et al. \cite{bally}).
The observations were made during orbits 313 (S(0) and S(3) lines),
 and 467 (S(1) and S(2) lines).
The wavelength bands were scanned in the   SWS02
mode with a typical on-target time of 100 s. 
The spectral resolution ($\lambda/\Delta\lambda $) 
of this mode is $\sim$ 1000-2000 corresponding to
a velocity resolution of $\sim$ 150-300 \kms.
All the lines have broader profiles than those expected
for a point source by a factor 1.3-1.5, 
indicating that the sources are extended in the direction
perpendicular to the slit 
(see Valentijn \& Van der Werf \cite{valentijn99}). 
 The flux calibration is believed to be accurate to 30$\%$, 20$\%$, 25$\%$,
and 25$\%$ for the S(0), S(1), S(2), and S(3) lines, respectively (Salama
et al. 1997).
Data reduction was carried out with version  6 of the
 SWS Interactive Analysis at the ISO Spectrometer
Data Center at MPE.
Further analysis has been made using the ISAP
    \footnote {The ISO Spectral Analysis Package (ISAP) is a joint
    development by the LWS and SWS
    Instrument Teams and Data Centers. Contributing institutes are
    CESR, IAS, IPAC, MPE, RAL and SRON.}
software package. All lines have  been rebinned to one fifth of the
spectral resolution of the instrument.
 Fig. \ref{figobs}a-b shows the spectra,  and
the observed parameters  are given in Table 1.
The errors in the radial velocities of the \h2 lines listed in
this table have been estimated from the Gaussian fits.
The wavelength calibration uncertainties, expressed in
velocities, are typically of
20-40 km s$^{-1}$ for $\lambda > 12 \mu$m and
$\sim$ 25-60 km s$^{-1}$ for $\lambda < 12 \mu$m 
(Valentijn et al. \cite{valentijn96}). Thus, the calibration
uncertainties usually dominate the global error in the radial
velocities.
When one takes into account the errors from the Gaussian fits
and the wavelength calibration uncertainties, 
the central velocities of the \h2 lines
are in agreement with those measured from  the CO lines (section 2.2).
It is noteworthy that, the higher the signal-to-noise ratio
of the \h2 lines (S(1) lines), the better the
agreement  of the H$_2$ radial velocities with those of CO.

 We also present {\em Long Wavelength Spectrometer}
 (LWS; Clegg et al. \cite{clegg}; Swinyard et al. 
\cite{swinyard}) observations of these sources in grating mode (43-196.7 $\mu$m,
$\lambda$/$\Delta\lambda$~$\sim$~200). 
Fig. \ref{figobs} e-f  shows the LWS spectra. The 
spectral resolution was 0.29 $\mu$m for the 43-93 $\mu$m range and 0.6 $\mu$m
for the 80-196 $\mu$m range. The LWS aperture was 
$\sim 80^"\times 80^"$.       
The roll angle, which gives the orientation of the apertures,
was  90\degr $\pm$ 2\degr for both the SWS and the LWS observations.
Data were taken during orbits      315 and 318
and processed through the LWS Pipeline Version~7.
 The  individual detector scans were calibrated to within 10$\%$ of
each other, based on overlapping detectors. 
Post-pipeline analysis (including shifting the different detectors using  
dark currents   and defringing) was performed with ISAP.

\subsection{IRAM 30-m observations}
The J=1--0 line of $^{13}$CO and C$^{18}$O and the
J=2--1 line of C$^{18}$O were       observed  simultaneously with
the IRAM 30-m telescope  (Pico Veleta, Spain) in May  1997.
We used two SIS receivers at 3 and 1.3 mm
connected to two $512 \times 1$ MHz channel filter banks.
This configuration  provided a velocity resolution of
2.7  and 1.4 \kms     for the J=1--0 and J=2--1 lines respectively.
Typical system
temperatures were $\sim$~250 K for the J=1--0 line and  $\sim$~500 K for 
the J=2--1 line. The receivers were tuned to single side band with
rejections always larger than 10 dB that were checked against
standard calibration sources.
The beam size of the 30-m telescope was  22$^"$ and 11$^"$ at 3
and 1.3 mm respectively. Pointing
and focus were monitored regularly. Pointing corrections were always
found to be smaller than 3$^"$. Calibration of the data was made by
observing a hot and cold loads with known temperatures,
and the line intensities were converted to main beam
brightness temperature, $T_{\rm MB}$, using main beam efficiencies of 
0.74 and 0.48 at 3 and 1.3 mm respectively.
The spectra  are  shown in Fig \ref{figobs}c-d and the observed parameters
as derived from Gaussian fits are listed in Table 2.       
 
\section{Analysis}

\begin{figure*}[p]
\centerline{\psfig{figure=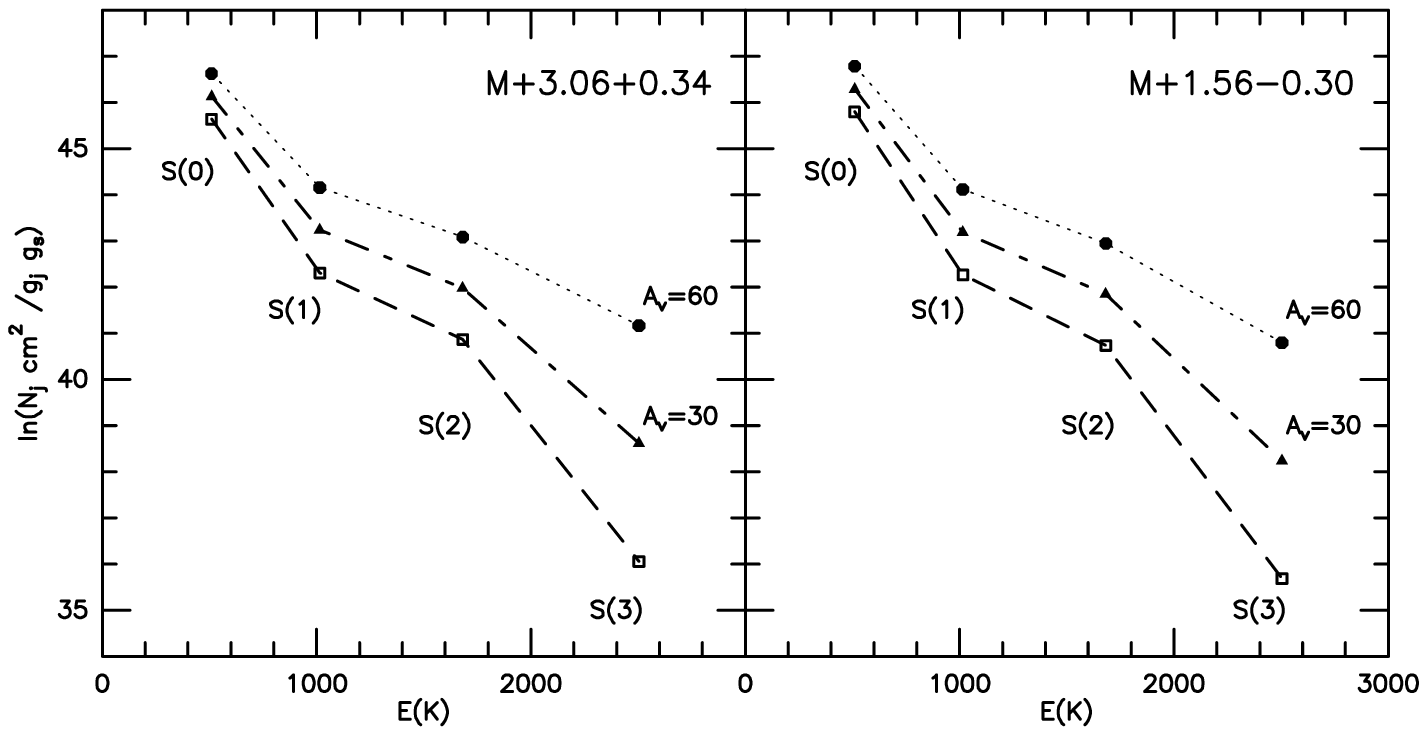,bbllx=83pt,bblly=510pt,bburx=518pt,bbury=758pt}}
\caption[]{Rotational plots.
The results are displayed for three different values
of the visual extinction: 0, 30, and 60 mag. We can see the typical
zigzag distribution of a non-LTE OTPR. Extinctions higher than 60 mag are 
 needed to have the smooth characteristic  curve of emission
 arising from gas with an equilibrium OTPR and a temperature gradient.}

\label{figext}
\end{figure*}

In Fig.  \ref{figext}  we show  the  \h2  rotational 
diagrams for the two sources.
The open squares correspond to the column densities as measured with
ISO, without any correction for the different apertures in       the
different lines and for the dust extinction. The rotational diagrams 
for the two sources  show
a zig-zag distribution  since  the column densities in the ortho-\h2
levels J=3 and 5 are  lower than those expected from the para-\h2  levels
for the LTE OTPR.
 For the typical
temperatures involved in these transitions ($\sim$~200 K), 
the LTE OTPR is   $\sim$~3.
Similar  rotational diagrams derived from  the \h2 pure-rotational lines
have been previously found in HH54 
by Neufeld et al. (\cite{neufeld}) and in NGC7023 by Fuente et al. 
(\cite{fuente99}). 
For these sources where extinction is known to be low, the immediate
conclusion was that the OTPR was not in  LTE.

The \h2 emission has been detected in all
sources of our sample indicating that the  \h2 emission in the GC must 
be relatively widespread and extended 
(Mart\'{\i}n-Pintado et al. \cite{mp99b}). 
This is also suggested from the measured linewidths of the
\h2 lines (see Section 2).
Anyhow, even in the extreme case that the \h2  emission were point-like,
the corrections for the different apertures would be small
and would not affect substantially the conclusions about the OTPR.
For a  point-like source    the S(0) line will be more
diluted than the S(1) and S(2) lines  because of the larger beam
($20^" \times 27^" $ instead of $14^" \times 27^" $). 
The situation for the S(3) line will be the opposite since
the aperture  at this wavelength is $14^" \times 20^"$.
Therefore, in this limit case, the column densities in the level
J=2 (derived from the S(0) line) averaged in a beam of $14^" \times 27^"$
would be larger by a factor of 1.4, while on the opposite,
the beam-averaged column density in the J=5 level would be
smaller by a factor of  1.4. Hence, the
correction for different apertures, cannot explain
the zig-zag distribution in the rotational diagram.

A more critical correction is that for  the extinction 
produced by the foreground material. As described by
Martin-Pintado et al. (\cite{mp99b}) the weakness of the S(3) line  in the GC
clouds should be due to the extinction produced by the silicate 
feature at 9.7 $\mu$m in the foreground dust clouds.
In clouds with a LTE  \h2 OTPR, one can use the intensity of the S(3) line
to estimate the visual extinction once the  relative value for the
opacity at 9.7 $\mu$m to the 0.55 $\mu$m opacity is 
known.  
One could, in principle,  apply corrections for increasing extinctions  until 
the column density in the  J=5  level is consistent with  the 
column  densities derived for other levels, i.e., until
the rotational plot is a straight line  
(in the case of a Boltzmann distribution  with one source temperature) or
a smooth curve (in the case of a temperature gradient). 
In clouds with a non-equilibrium OTPR one could use
an equivalent method using only ortho-\h2 levels, but obviously
more than two levels are     needed.
The effect of foreground extinction on the rotational diagram
 is  illustrated in
Fig. \ref{figext}, where the observed fluxes have been corrected for 30 (filled
triangles) and 60 mag  (filled circles) of visual extinction, using the 
extinction law of Draine \& Lee (\cite{draine}). 
Visual  extinctions larger than 60 mag
are needed  for      consistency between  the S(1) and S(3) line intensities
and  a LTE OTPR.  In this case, the curvature of the rotational plots
suggests   the presence of a large temperature gradient in the \h2 
emitting region.
To constrain the visual extinction toward these sources, 
in the following sections we will estimate  the total column densities 
of dust and gas  from
measurements of the continuum dust emission, \m13co, and  \c18o
with a similar resolution to that of the SWS aperture.

\subsection{\h2  column densities from \c18o and \m13co observations}
We applied  the Large Velocity Gradient (LVG) approximation to 
our data, to derive       
the physical conditions and the column densities 
of molecular gas from the emission of
the J=2-1 line of \c18o and the J=1-0 lines of \c18o and \m13co.
The lines toward the two sources show complex profiles with two 
velocity components. 
From the line intensity
ratios one can see that these components have slightly different
physical conditions.
The \c18o J=2-1 to J=1-0 line ratio is  1.0-1.4 in \mtres
and cannot be determined for the other source. The J=1-0 \m13co to \c18o
ratio ranges between 5 and 14 in \mtres and is $>$~7 in M+1.56$-$0.30. 
To within a factor of 2, these values
are in agreement with the typical isotopic abundances 
found in the GC  for carbon and oxygen
(see Wilson \& Matteucci 1994) indicating that the
\m13co lines are optically thin.  
From the \c18o J=2-1 to J=1-0 ratio we derive for \mtres the \h2 
densities given in Table 3 for two cases: high kinetic temperature 
($T_{\rm K}$=100~K)
and low kinetic temperature ($T_{\rm K}$=20~K). For those \h2 densities
we have constrained  the total column densities using the 
\m13co line intensities.
When the \c18o lines were not detected 
the range of possible \m13co(1-0) column densities was obtained by
changing the \h2 density  between 10$^{3}$  and 10$^{4}$ \cmmt
for $T_{\rm K}$=100~K 
and between 10$^{3.5}$ and 10$^{4.5}$ \cmmt for $T_{\rm K}$= 20 K 
(see H\"uttemeister et al. \cite{huttemeister98}).
In the case of cold gas  and even higher  \h2 densities, the \m13co column 
densities will increase only in a factor of  1.3 since for low temperatures
and densities $\ge$ 10$^{4}$ the J=1--0 transition of \m13co is thermalized.  
The \h2 column densities, \Nh2,  in Table 3  have
been derived  from the \m13co column density and a fractional abundance
with respect to \h2 of 5~10$^{-6}$. They are  
typically of a few 10$^{22}$ cm$^{-2}$,
in good agreement with the values given by  
H\"uttemeister et al. (\cite{huttemeister98}). 
With these column densities, we have derived the total visual extinction,
$A_{\rm v}$, using the standard conversion factor:      
\Nh2(\cmmd)=$A_{\rm v}({\rm mag})\times 10^{21}$. Thus
the extinctions toward  the two GC sources studied   in this
paper are typically of  15-20 magnitudes.

\subsection{Dust column densities and temperatures}

From the LWS data we can make a direct estimate of  the
dust temperature and the dust  column densities toward both   sources.
 Though  the aperture of
the LWS is larger than that of the SWS, the dust emission in  the GC is
relatively smooth (Odenwald \&  Fazio \cite{odenwald}) and one does not
expect large variations within the LWS aperture.
The spectra for the two   sources have very similar shapes
with the maximum of the emission  at $\sim$ 100~$\mu$m, indicating
that the bulk of
the dust is relatively cold with temperatures below 30 K, in agreement  with
previous estimates (Odenwald \& Fazio \cite{odenwald}; Gautier et al. 
\cite{gautier}).

\begin{figure*}[p]
\centerline{\psfig{figure=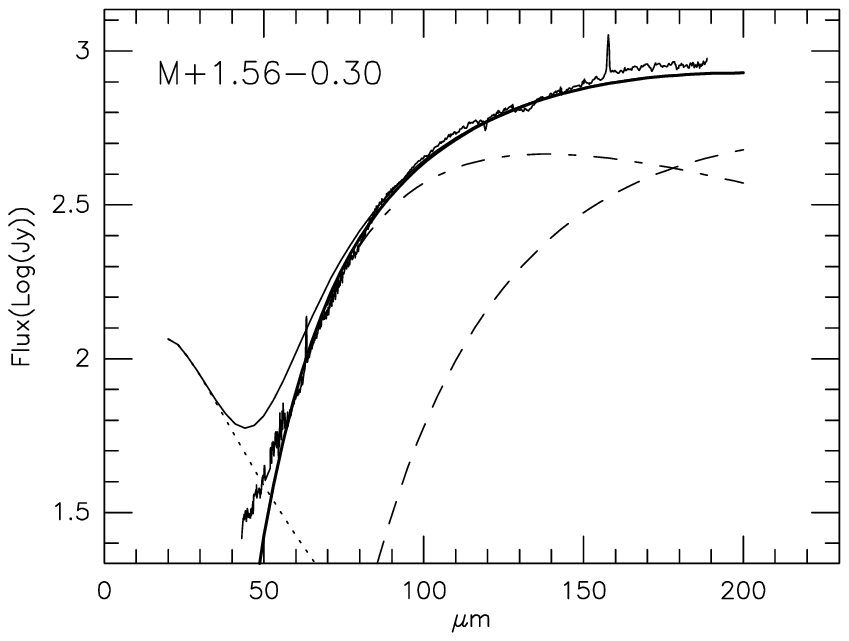,bbllx=80pt,bblly=501pt,bburx=378pt,bbury=739pt}}
\caption[]{The LWS spectrum of \muno. The thick solid line is the best fit
with two components of temperatures 15 K (dashed line) and 27 K (dot-dashed
line). The solid line is the total emission for an equivalent column density
of \Nh2$=5\times 10^{18}$ cm$^{-2}$ of hot (250 K) dust with
$\Omega = 20^"\times 20^"$ (dotted line) located behind the cold dust.
We have assumed that 
the hot component is extincted by the cold component.}
\label{figdust}
\end{figure*}

The  data cannot be fitted with only one gray body. For simplicity,
 we have considered a model  with two gray bodies of   temperatures
$T_1$ and $T_2$. The total flux, $S_\lambda$, is    given by:
\begin{equation}  
\label{bb}
  S_\lambda=\Omega [B(T_1,\lambda)(1-e^{- (1-f) \tau(\lambda)})
     +B(T_2,\lambda)(1-e^{-f \tau(\lambda)})]
\end{equation}  
\noindent    where $\Omega$ is the solid angle of the continuum source, 
$B(T)$ is the Planck function,  $f$ is the fraction of the  opacity 
due to the warmer component ($T_2$), and 
$\tau(\lambda)$ is the total opacity at wavelength  $\lambda$. 
In this model,  the ratio of the visual extinction, $A_{\rm v}$, to
the total optical depth at 30 $\mu$m is taken from the
 Draine \& Lee (\cite{draine})  extinction law and    the opacity
for  $\lambda$  $>$ 30 is      given by:          
\begin{equation}
\label{tau}
   \tau(\lambda)=0.014 A_{\rm v}(30\mu{\rm m}/\lambda)^{\alpha}
\end{equation}
where $\alpha$ is the spectral index of the dust emission.
In accordance  with previous estimates for the envelope of Sgr B2
 (Mart\I n-Pintado et al. \cite{mp90}) and for the GC background of the cold
core GCM 0.25+0.11 (Lis \&  Menten \cite{lis}), 
we have taken $\alpha \simeq 1$.
We have assumed extended emission
($\Omega=\Omega_{\rm LWS}$) and
then we have fitted the continuum spectrum  
with  $f$, $A_{\rm v}$, $T_1$, and $T_2$  as free parameters. 
As an example,  we show in Fig. \ref{figdust}  the best fit to the LWS
spectra towards  \muno obtained  with $A_{\rm v}$=40, $T_1$=15~K, 
$T_2$=27~K and $f$=0.1.
 
Table 4 lists the results of the parameters for the best fits for the
two sources. 
The visual extinctions derived for the two sources are 30 and 40 mag.
These values are in agreement, to  within a factor of 2, 
with those derived from the CO data. 

The dust emission is dominated by the cool ($T \sim$ 15 K) component 
($\tau_{\rm v_1}\sim (1-f) \tau_{\rm v}$), 
while the slightly warmer component ($T \sim$ 30 K) contributes only 
10$\%$-20$\%$ to the total optical depth ($\tau_{\rm v_2}\sim f \tau_{\rm v}$).
We can also fit the spectra with larger spectral indexes by
increasing the dust column densities.
For instance, an  spectral index of 1.5 will increase the visual extinction
to 50-100 mag.
These high values of $A_{\rm v}$ are very unlikely since they are 
almost one order of magnitude higher
than the estimates   made from CO (see Table 3).  

Since the extinction derived from CO and the continuum accounts for the
total gas and dust along the line of sight, they must represent  an
{\em upper} limit to the extinction to the \h2 emitting region.
Considering     the uncertainties introduced by the unknown spectral index
and the many free parameters 
in the dust column density determination,
in the following discussion we will assume  that upper limits to the
 visual extinction
of the \h2 emitting region are those derived from the CO emission, namely,  
16 magnitudes for \mtres and 20 mag for M+1.56$-$0.30. These values 
are within a factor of two of estimates  obtained   from the  
total dust column density.


\subsection{Warm \h2: ortho-to-para ratio  and column densities.}

\begin{figure*}[p]
\centerline{\psfig{figure=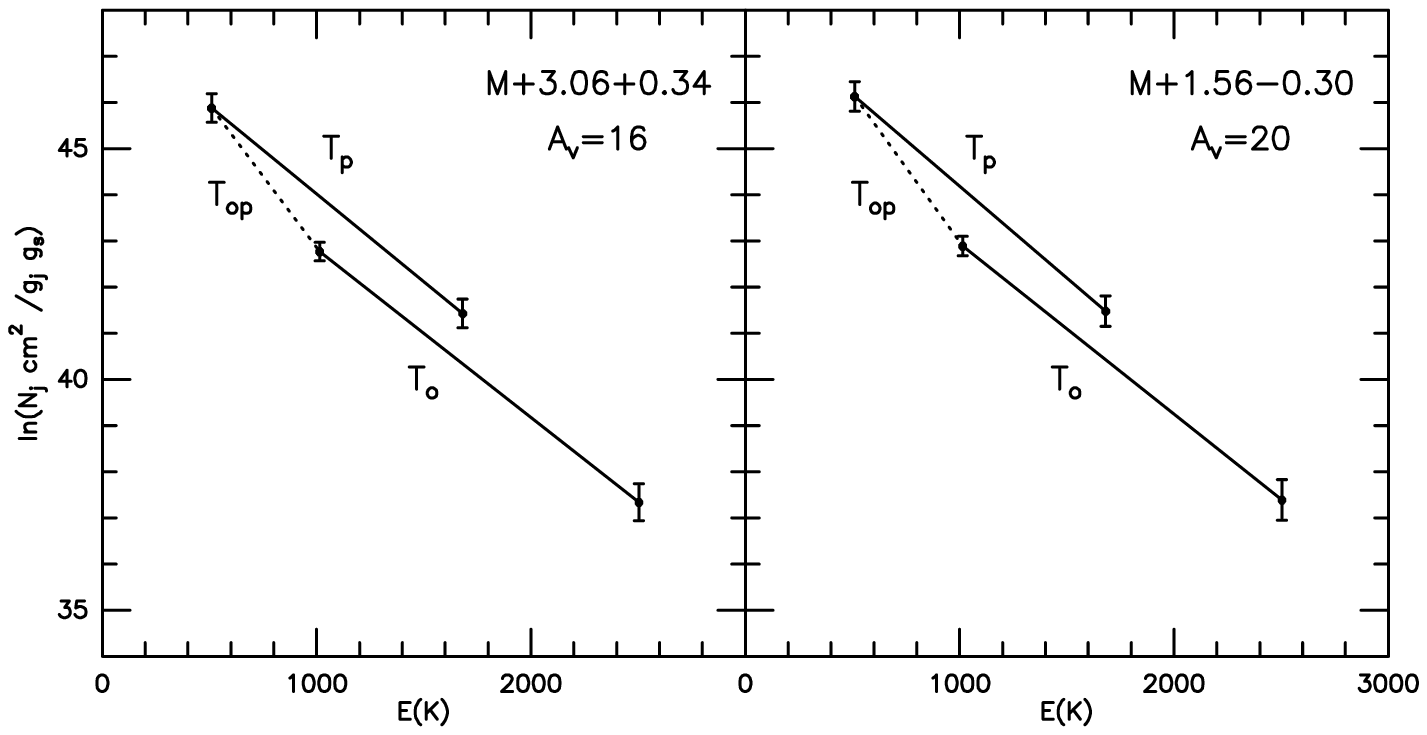,bbllx=85pt,bblly=494pt,bburx=525pt,bbury=772pt}}
\caption[]{Rotational plots after correcting for the most probable 
extinctions. The slope of the lines is  proportional to the inverse
of the temperature. $T_{\rm p}$ is the rotational temperature between
the para-\h2 levels,  $T_{\rm o}$ between the ortho-\h2 levels, and
T$_{op}$ between the ortho-level J=3 and the para-level J=2. Error bars
take into account the calibration uncertainties and the errors in
the Gaussian fits.}
\label{figotpr}
\end{figure*}

As discussed at the beginning of Sect. 3        
 the \h2 OTPR depends on the correction for
extinction. In the previous sections we have  
estimated the extinction for the two
clouds and  Fig \ref{figotpr} shows the \h2 rotational  diagrams for \mtres
 and \muno corrected for  the estimated extinctions. The error bars 
take into account the  errors in the Gaussian fits of the lines
and the calibration uncertainties. From these data, 
we  derive an {\em ortho rotational temperature}, $T_{\rm o}$, from 
the ortho-\h2 levels J=3 and J=5. In the same way, one can 
define a {\em  para rotational  temperature}, $T_{\rm p}$, 
 derived from the para-\h2 levels J=2 and J=4, and an 
{\em ortho-para temperature}, $T_{\rm op}$, 
derived from   the ortho level J=3 and the para level J=2.
These temperatures are listed in Table 5.  
As we see, $T_{\rm p}$  is $\sim$~250~K for both sources while  $T_{\rm o}$   
   is slightly
higher ($\sim$~270~K) indicating  the presence of a  moderate 
temperature gradient. 
This effect is more definite  in other sources of our sample, where the
S(4) and S(5) lines, which trace clearly higher temperatures,
have also been observed (Mart\I n-Pintado et al. \cite{mp99b}).    
For the present sample, $T_{\rm op}$ is $\sim 160$ K, much smaller 
than $T_{\rm p}$ and $T_{\rm o}$  indicating a non-LTE 
OTPR. In terms of these temperatures, the OTPR measured from our data
will be given by:
\begin{equation}
\label{otpr}
{\rm OTPR}={\rm OTPR}_{\rm LTE}(T_{\rm p} )\exp (\frac{1}{T_{\rm p}}-
                        \frac{1}{T_{\rm op}})
\end{equation}
where OTPR$_{\rm LTE}$(T) is the LTE OTPR at    temperature $T$. 
As mentioned before,   OTPR$_{\rm LTE}$ is $\sim$~3 for $T \ge$ 200 K. 
Using Eq. (\ref{otpr}), one finds an OTPR of $\sim$~1
for both sources (see Table 5). 
Increasing the extinction will make the \h2 OTPR closer to the equilibrium
value, however extinctions $>$ 70 mag will be
required to give an LTE  OTPR ratio. 
Such  large visual
extinctions  are very unlikely from the molecular line and continuum data
discussed in the previous sections. We therefore conclude 
that for the two sources
the \h2 OTPR is {\em not}  in equilibrium.
Since the   estimated  error is $\sim$ 0.4, 
we can take  $\sim$ 1.4 as a conservative upper limit for the OTPR in
these two sources.                                

Extrapolating the populations in the  J=2 and J=3 levels  to
the  J=0 and J=1 levels , respectively, as two 
different species at temperature $T_{\rm p}$, one finds  that the total
column densities of warm \h2 are $\sim$ 2~10$^{21}$ \cmmd. This must
be considered as a {\em lower} limit to the 
actual warm \h2 column density since the 
populations of the lowest levels (J=0 and J=1)
 can be increased significantly by  colder, 
though still warm ($\sim$ 100 K) gas. Of course, if extinction
is higher column densities will  also increase. This implies that
the measured  ratio of warm \h2 to cold gas traced by CO is 
at least 15$\%$.

High gas kinetic temperatures in these two clouds are known to be present
from  the \NH3 observations of  H\"uttemeister et al. 
(\cite{huttemeister93}). The rotational
temperatures derived from the (4,4) and 
the (5,5) metastable inversion
lines of \NH3  are in good agreement with the 
temperatures derived  in this paper using the lowest \h2 pure-rotational lines.
Extrapolating the populations in the (4,4) and the (5,5) \NH3 
levels to  lower levels with the  rotational temperature derived
for each source by H\"uttemeister et al. (\cite{huttemeister93})  one finds 
a column density of warm \NH3 of   $\sim 7~10^{14}$ \cmmd 
 in both sources. Taking into account the  warm \h2 column densities
given above, we
find a \NH3 abundance of  (2-4)~10$^{-7}$, similar to the value
obtained by Mart\I n-Pintado et al. (\cite{mp99a}) in the expanding shells 
of the envelope of Sgr B2. A similar    abundance is also 
obtained when we compare 
the column densities of cold ($\sim $ 20 K) \NH3  
(H\"uttemeister et al. \cite{huttemeister93})
 and the \h2 column densities derived by our \m13co 
and \c18o data.

\subsection{Warm dust column densities}
If the gas and dust are coupled, one expects 
that  the dust associated with the warm \h2 component
would be an intense continuum emitter  in the mid- and far-IR.
There is no hint of such dust component in our  data,
as      shown in  Fig.~\ref{figdust}, where we
represent (as a dotted line) the emission of a gray body with 
a temperature of 250~K and the size of the SWS aperture 
attenuated by the total column density of the cold component. 
The equivalent \h2 column 
density of warm  dust used to simulate  the emission in
Fig. \ref{figdust} is   only 5~10$^{18}$ {\cmmd}.
Even this small column density should have been detected.
 Hence, we
can rule out a dust component coupled to the warm gas 
with a column density larger than 2~10$^{-3}$ times that of the warm \h2.
On the other hand, the comparison
of CO emission with the cold dust emission shows agreement with the standard 
gas-to-dust ratio within a factor of two.

\section{Discussion}

\subsection{Heating of the warm component}
The large column densities of warm  \h2 and the low column densities
of associated warm dust require
a heating mechanism that heats selectively the gas maintaining
the dust at much lower temperatures. A PDR with  an incident 
far-ultraviolet (FUV) flux $G_0$ of $\sim$ 100 (measured in units of
1.6~10$^{-3}$ ergs \cmmd s$^{-1}$) can heat 
the gas via photoelectric effect  in the grains 
to temperatures of 100-200 K in the external layers of the cloud 
without heating the dust to temperatures
above 30 K (see Hollenbach et al. \cite{hollenbach}). 
However, the large gas phase \NH3 abundance, as  derived in Sect.  3.3,  
is not possible            in such a PDR scenario. 
The evaporation temperature of  \NH3 is $\sim$ 75 K, therefore it cannot 
be evaporated from grain mantles at only 30 K. Even in the case 
that evaporation occurs, the UV   radiation that heats 
the dust would destroy the fragile \NH3 molecule. This is the behavior
 found in NGC 7023 where the \NH3 abundance is   $\sim 10^{-8}$ in 
the well shielded region and decreases  by more than a factor 
of 30 towards the  region where the UV radiation increases and 
the dust temperature is $\sim$ 70 K   (Fuente et al. \cite{fuente90}).

Shocks have been invoked  as an important heating 
mechanism for the GC clouds
(Wilson et al. \cite{wilson82}; Mart\I n-Pintado et al. \cite{mp97}; 
H\"uttemeister et al. \cite{huttemeister98}). In fact,
 \muno belongs to the ``$l=$1\fdg5-complex", where H\"uttemeister et al. 
(\cite{huttemeister98}) derived the highest SiO abundance
within their sample, while the CS abundance (which
traces all dense gas, not just the part that has been subjected
to shocks)  is not enhanced.               
They interpreted  the SiO enhancement  to be produced by                 
 large scale dynamic effects, 
proposing that in this      complex,  gas sprayed from the 
intersection of the $x_1$ and $x_2$ orbits is crashing into material 
that is still on $x_1$ orbits in the context of a bar morphology.
In our sample, \muno is also the source with the highest SiO to CS ratio.
Furthermore, Dahmen et al. (\cite{dahmen}),  studying the HNCO emission in 
this region, found evidence for collisional excitation by shocks.

On the other hand, \mtres is located close to one of the CS cores detected by
Stark \& Bania (\cite{stark}) in the  ``Clump 2" complex. These dense cores are
gravitationally bound but most of the CO is emitted from the
lower density gas, not bound to 
the cores. Stark \& Bania (\cite{stark}) suggested  that this material 
is the result of   tidal stripping of the cores. 
It is definite that shocks can play a role to explain the 
large column densities of warm \h2 and the relatively large abundances of
\NH3, as well as the high kinetic temperatures in these clouds. In addition,
transient heating by shock waves provides a natural explanation for
\h2 OTPRs out of equilibrium. 
 
\begin{figure*}[p]
\centerline{\psfig{figure=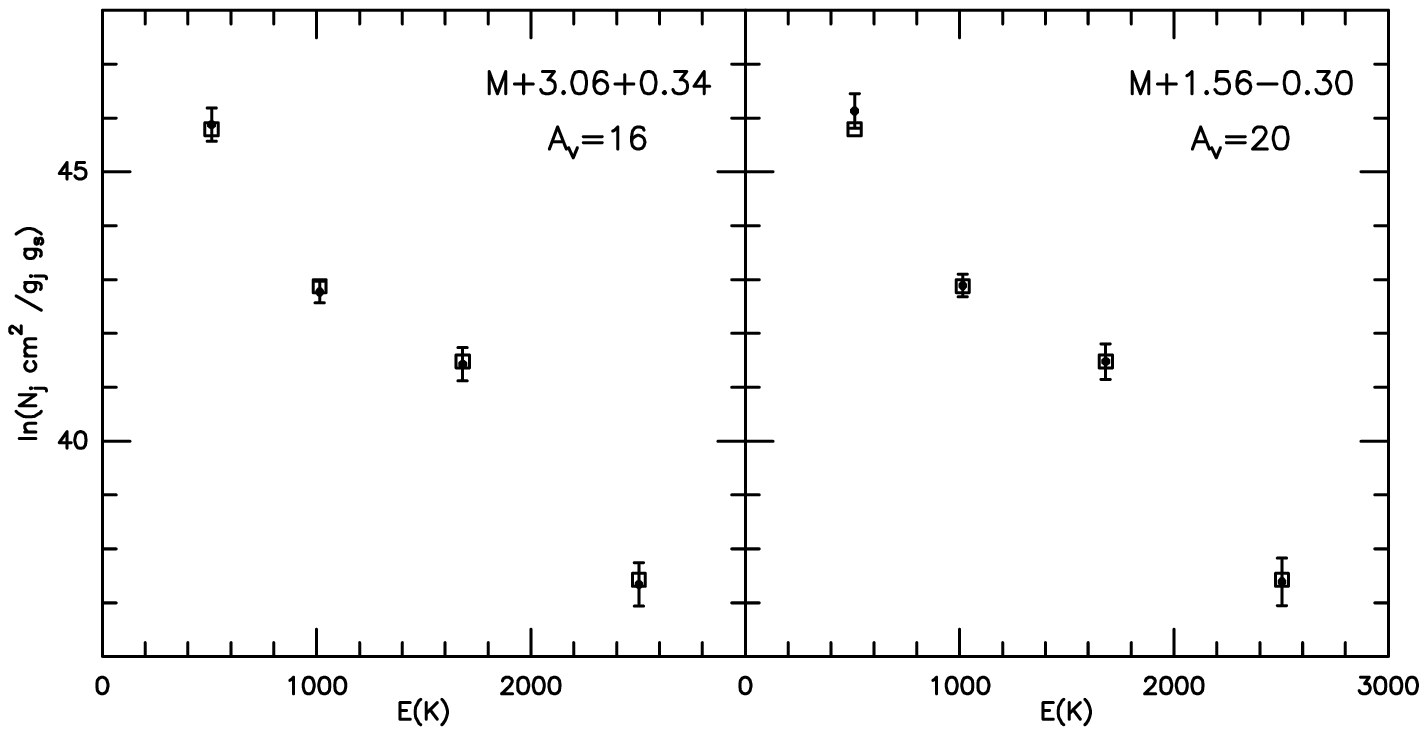,bbllx=77pt,bblly=503pt,bburx=527pt,bbury=770pt}}
\caption[]{Filled circles and error bars as in Fig. 4
Empty squares are the expected column densities using the model of
Timmermann (1998) with preshock density $\sim 2~10^5$ \cmmt,
shock velocity 10 \kms, and preshock OTPR=1}
\label{figtim}
\end{figure*}

 We have compared the results of the model calculations for slow shocks
by Timmermann (\cite{timmermann}) with our \h2 data. Interpolating the \h2 line
strengths predicted by the model for a preshock OTPR=1 as a function of density,
we  found that a  shock with velocity of 
10 \kms and a preshock \h2 density of  $\sim 2~10^5$ \cmmt reproduces
the observed line intensities. 
 The results are displayed in Fig. \ref{figtim} 
in the form of rotational plots. Open squares are the predicted column
densities,
while filled circles are the  values derived from observations after
correcting for extinction. 
Though the observed flux in the S(0) seems to be slightly
larger  than   in the model, the agreement is excellent, and
calibration errors can account for the discrepancies.
The preshock density seems somewhat high but it is plausible since
the S(3) line is {\em apparently}  thermalized, which implies a lower limit
to the \h2 density of $\sim 10^4$ \cmmt. 
In any event, the widespread distribution of the HCN emission (Jackson et al.
\cite{jackson}) shows      that densities of $\sim 10^5$~\cmmt are
common in the GC.

\subsection{The ortho-to-para ratio}
The main    processes that  affect the OTPR of H$_2$ are     
proton exchange collisions with H$^+$ and
reactive collisions with H atoms. Ortho-para conversion in grain surfaces
is  thought  to be less efficient. 
The rate coefficient for the proton exchange reaction 
\begin{equation}
\label{conv}
{\rm H}_2({\rm ortho})+{\rm H}^+ \rightleftharpoons {\rm H}_2 ({\rm para})
 + {\rm H}^+ + 170.5~{\rm K}
\end{equation} 
is $\sim 3~10^{-10}$ cm$^3$ s$^{-1}$ (Gerlich \cite{gerlich}).
The analogous reactions with H$_3^+$ and H$_3$O$^+$ may also occur  
at a similar rate (see e.g. Le Bourlot et al. \cite{lebourlot}).
This rate gives an ortho-para conversion timescale, $\tau_{\rm conv}$,
of $\sim 100/n(^+)$ yr, where $n(^+)$ represent the density of
H$^+$, H$_3^+$ or H$_3$O$^+$  in \cmmt.
One should note that the actual conversion time can be
a factor of 10 larger than $\tau_{\rm conv}$ (see Flower and Watt 1984).

The rate coefficient for the reactive collisions with H atoms
\begin{equation}
\label{convh}
{\rm H}_2({\rm ortho})+{\rm H} \rightleftharpoons {\rm H}_2 ({\rm para})
 + {\rm H} + 170.5~{\rm K}
\end{equation}
is $\sim 8~10^{-11} {\rm e}^{-(3900/T)}$ cm$^3$ s$^{-1}$ 
(see e.g. Le Bourlot et al. \cite{lebourlot}).
Due to the high activation barrier of this reaction (3900 K),
in cold and dense molecular clouds the dominant process will be
proton exchange collisions. This is also true for low velocity shocks
of $\sim$ 10 \kms since the maximum temperature achieved in the post-shock
region is only $\sim 300$ K. 
If the H$^+$ and H$_3^+$ densities  ($n$(H$^+$), $n$(H$_3^+$))  
in the postshock region
of a 10 \kms shock were as high as $\sim 10^{-3}$ \cmmt 
(see Timmermann \cite{timmermann}), $\tau_{\rm conv}$                        
would be $\sim 10^{5}$ yr.
It is worth-noting that recent models    for ortho-para conversion
in shocks by Wilgenbus et al. (\cite{wilgenbus}) find much lower
H$^+$ and H$_3^+$ densities in the postshock region.
In this case, the timescale for ortho-para conversion would be $> 10^5$ yr.
On the other hand, the time needed for the passage of  the proposed 
10 \kms velocity shock,
from the point where the neutral gas starts to heat up to
the point where the gas has reached interstellar temperatures again,
is  $\lsim 10^{4}$~yr (see Timmermann \cite{timmermann}). 
However  the timescales in which the neutral gas is at high
temperatures are much shorter. Hence, if the initial OTPR was lower
than 3, the heating-cooling of the gas is too fast for the
OTPR to reach the equilibrium at the  temperatures of the shocked
material.

Shocks with velocities $ >$ 20 \kms 
heat the gas to temperatures  $ >$ 700 K. 
Then,  collisions
with H would be the main ortho-to-para conversion mechanism, and indeed,
the ortho-para conversion timescale would be low enough to  obtain  
at least some conversion in the
shock timescale as in the source HH54 (Neufeld et al. \cite{neufeld}).
However,  the lines ratios in \muno and \mtres
cannot be explained with a preshock OTPR of $<$ 1 and a shock
with velocity $>$10 km s$^{-1}$. 
Therefore, the observed OTPR in these clouds must be
approximately  the preshock OTPR. 
 This conclusion is independent of any shock model since
the low temperatures involved by a 10 \kms shock are not sufficient for the
H$_2$-H reactive collisions to be effective and, even for the largest
predicted H$^+$ and H$_3^+$ abundances, the proton exchange reactions
are not fast enough to give ortho-para conversion in the shock timescale.

If the OTPR of the preshock gas was in equilibrium at the gas temperature, 
the  temperature should be $\sim$ 80 K. In this case, the preshock
gas should have been already heated
before the shock front compresses and heats the gas to 250~K.
However, there is no strong reason to believe that the preshock OTPR
should be in equilibrium at the preshock temperature. 
The \h2 molecule is formed  mainly on the grain surfaces by
a highly exothermic reaction. Thus, if it is  rapidly ejected 
to gas phase the OTPR will  be the typical 
OTPR at high temperature, i.e., 3.
On the other hand, if it is not evaporated immediately from
the grain there will be ortho-to-para conversion by collisions
with radicals, impurities or defects and the OTPR could reach
the equilibrium value at  $\sim$ 30 K (dust  temperature) of $\sim $0.01.
In our case, the preshock OTPR of  $\sim 1$ suggests that the \h2 molecules
were      ejected from the grains with  OTPR $>$1.
Afterwards, 
this ratio  could decrease due to proton exchange processes.

The equilibrium proton abundance in dense ($n$(H$_2$) $\sim 10^5$ \cmmt)
clouds, where photoprocesses are not important, depends 
mainly on the ionization by cosmic rays and 
on charge exchange reactions with neutral molecules.
Modeling the chemistry  of dense PDRs, 
Sternberg and Dalgarno (\cite{sternberg2}) found $n$(H$^+$) 
of $\sim 10^{-5}$ \cmmt in the well UV-shielded region 
for a cosmic ray ionization rate ($\zeta$)
of 5 $10^{-17} {\rm s}^{-1}$, implying $\tau_{\rm conv} \sim 10^7$ yr.
A similar timescale
is obtained for proton exchange collisions with H$_3^+$. 
The density of H$_3$O$^+$ could reach $10^{-4}$ cm$^{-3}$ and thus
$\tau_{\rm conv}$  could decrease by a factor of 10.
Nevertheless, the actual time to reach the LTE OTPR would be longer.
 Flower \&  Watt (1984) have studied the temporal evolution of the
OTPR in molecular clouds.          
Using the same rate coefficient as above 
for the proton exchange process,  they have shown that for  H$^+$ densities 
         \footnote{In their model these proton densities were 
                  obtained with  $\zeta =10^{-17} - 10^{-18} {\rm s}^{-1}$ 
                    using a simplified chemical network. }
of $10^{-4}-10^{-5}$ \cmmt
the actual time needed for an  OTPR=3 to  be in equilibrium at 30 K 
(similar to  the observed cold component in the GC 
clouds) is $\sim 10^7 -10^8$ yr. 
In particular, if $n({\rm H}^+)$ (or $n({\rm H}_3{\rm O}^+)$) is 
$\sim 10^{-4}$ \cmmt, then   $\sim$ 5 10$^6$ yr will be needed 
to have an OTPR=1.
Assuming that the \h2 was ejected to gas phase after formation
with an OTPR \lsim 3,
the shock front reached the cloud  approximately
10$^6$ yr after the formation of the \h2 molecules,
since this is the time needed for an OTPR$\sim$3 to descend to $\sim$1
in a dense molecular cloud.

\section{Conclusions}   

 We have presented ISO SWS observations of the S(0), S(1), S(2), and S(3) 
pure-rotational lines of \h2 and LWS observations of 
the dust continuum and IRAM-30m \m13co and \c18o observations  toward
 the GC molecular clouds \muno and \mtres. 
Using the CO data and dust column densities from the LWS spectra we estimate
$\sim$ 20 mag of visual extinction toward these sources. 
The two estimates   of the extinction agree within a factor
of 2 for the standard gas-to-dust conversion factor. 
According to   the two components scenario proposed by
H\"uttemeister et al. (\cite{huttemeister93}), the low-J CO 
emission  arises from the cold gas component and is 
coupled to the dust  at a temperature of $<$ 30 K. 
The warm component  ($T \sim$ 250 K) column density
observed directly in \h2 is, at least,  $\sim$ 15 $\%$ of the cold one and
 would have very little warm dust associated with it. 
From  the LWS spectra we set a conservative upper limit to
the warm gas column density associated with the warm dust 
 of 2~10$^{-3}$ times that of the warm \h2 column density.

After correcting for the dust extinction,
we  derive an OTPR of 1.0 $\pm$ 0.4, which is far from the LTE value
expected for the gas temperatures of 250 K.
We have also compared the warm \h2 column densities to the
\NH3 observations by H\"uttemeister et al. (\cite{huttemeister93}),
and derived \NH3 abundances of $\sim$ 2~10$^{-7}$, similar to those 
in the cold component (H\"uttemeister et al. \cite{huttemeister93}).

The low dust temperatures, the high \NH3 abundances, the large CO linewidths, 
the non-LTE \h2 OTPR, in addition to the high gas temperatures suggest that
the warm gas component is  heated  by  low velocity
shocks with speeds of $\sim$ 10 \kms. 
To explain the OTPR we propose the following scenario.
 \h2 is formed in the grain surfaces and ejected
to gas phase with OTPR\lsim 3. After $\sim 10^6$ years, the time required
to reach      the preshock OTPR=1,
a low velocity shock heated the gas to  the observed temperatures
of 250 K, but the OTPR was  almost unaltered because
the timescale for the passage of such a shock is  much
shorter than the ortho-to-para conversion timescale.
Taking into account the shock timescale this occurred less than
10$^4$ yr ago. 
It is interesting to note that the timescale of the cloud's
galactic rotation period is also $\sim 10^6$ years. This fact suggests
that the origin of the shocks can be related to large scale dynamics
of the GC region.

\begin{acknowledgements}
We thank S. Cabrit for her helpful comments on the ortho-to-para
conversion mechanisms.
We acknowledge support from the ISO Spectrometer Data Center at MPE,
funded by DARA under grant 50 QI 9402 3. NJR-F, JM-P, PdV, and AF have
been partially supported by the CYCIT and the PNIE under grants
PB96-104, 1FD97-1442 and ESP97-1490-E. 
NJR-F acknowledges {\it Conserjer\I a de Educaci\'on y Cultura de la
Comunidad de Madrid} for a pre-doctoral fellowship.
\end{acknowledgements} 
%

\begin{table*}[p]
\caption{Observational parameters of the \h2 lines. 
           Fluxes in units of 10$^{-20}$ W \cmmd. Heliocentric
         velocities ($v_{\rm lsr}$) in km s$^{-1}$. 
         Numbers in parenthesis are 1$\sigma$ errors of the
        last significant digit of the Gaussian fits.}
\begin{tabular}{lrrrrrrrr}
\hline
\multicolumn{1}{l}{} & 
\multicolumn{2}{c}{S(0)} & \multicolumn{2}{c}{S(1)   } & 
 \multicolumn{2}{c}{ S(2)   } & \multicolumn{2}{c}{S(3)  }  \\
\multicolumn{1}{l}{Wavelength} &
\multicolumn{2}{c}{28.2$\mu$m} & \multicolumn{2}{c}{17.0$\mu$m  } &
 \multicolumn{2}{c}{12.8$\mu$m}  & \multicolumn{2}{c}{9.7$\mu$m} \\
\multicolumn{1}{l}{Aperture } & 
\multicolumn{2}{c}{20$^"\times 27^"$ } & \multicolumn{2}{c}{14$^"\times 27^"$} &
\multicolumn{2}{c}{14$^"\times 27^"$}  & \multicolumn{2}{c}{14$^"\times 20^"$} \\
\multicolumn{1}{l}{} &
\multicolumn{1}{c}{Flux} & \multicolumn{1}{c}{$v_{\rm lsr}$} &
\multicolumn{1}{c}{Flux} & \multicolumn{1}{c}{$v_{\rm lsr}$} &
\multicolumn{1}{c}{Flux} & \multicolumn{1}{c}{$v_{\rm lsr}$} &
\multicolumn{1}{c}{Flux} & \multicolumn{1}{c}{$v_{\rm lsr}$} \\
\hline
 M+3.06+0.34   &6.9(6) & -13(7) &19.6(6) &  3(4)   &16(3)& -30 (20) 
                                 &1.6(5) &80(20)\\  
 M+1.56$-$0.30 &8.1(9) & -80(10)  &19(1)     & -43(4) & 14(3)& -50(20) 
                               &1.1(4) & -30(20)\\ 
\hline
\end{tabular}
\end{table*}


\begin{table*}[p]
\caption{Observational parameters of the \c18o and \m13co lines.
          Digits in parentheses are the errors in the last significant digits 
         (rms of the Gaussian fits). For \c18o(2--1) linewidths
          were fixed, when detected, to that of the \c18o(1--0) lines.}
\begin{tabular}{llllllllllll}
\hline
Source & Position   & \multicolumn{2}{c}{\c18o(1--0)$^a$}& &
\multicolumn{2}{c}{\c18o(2--1)$^a$} & &
\multicolumn{2}{c}{\m13co(1--0)} \\
\cline{3-4}\cline{6-7}\cline{9-10}
\multicolumn{1}{l}{      } &
\multicolumn{1}{l}{$v_{\rm lsr}^{^{13}{\rm CO}}$} &
\multicolumn{1}{c}{$\Delta v$ } &\multicolumn{1}{c}{$T_{\rm MB}$} & &
\multicolumn{1}{c}{$\Delta v$ } &\multicolumn{1}{c}{$T_{\rm MB}$} & &
\multicolumn{1}{c}{$\Delta v$ } &\multicolumn{1}{c}{$T_{\rm MB}$}  \\
\multicolumn{1}{l}{      } &
\multicolumn{1}{l}{km s$^{-1}$} &
\multicolumn{1}{c}{km s$^{-1}$ } &\multicolumn{1}{c}{K} & &
\multicolumn{1}{c}{km s$^{-1}$ } &\multicolumn{1}{c}{K} & &
\multicolumn{1}{c}{km s$^{-1}$ } &\multicolumn{1}{c}{K}  \\
\hline                                                         
M+3.06+0.34 &10.5(4) &6(1) &0.80(5)& &6(0) &0.8(2)& &10(1) &2.3(2) \\ 
            &27.9(6) &8(3) &0.44(5) & &8(0) &0.6(2)& &18(2) &2.6(2) \\
M+1.56$-$0.30 &-51(1) &--- & \le 0.14 &&--- & \le 0.20 &&19(5) &0.8(4) \\
            &-24.1(6) &--- & \le 0.14 &&--- & \le 0.20 &&21(1) &1.8(4) \\
\hline
\end{tabular}
    \begin{list}{}{}
       \item[$^a$] Limits are  3$\sigma$ assuming 
                           the same $\Delta v$ than that for the  $^{13}$CO.
    \end{list}
\end{table*}
 
\begin{table}[p]
\caption{Derived physical conditions for the 
\m13co and \c18o lines for $T_{\rm K}$=20 K and $T_{\rm K}$=100 K}    
\begin{tabular}{lrrrr}
\hline 
\multicolumn{1}{l}{Source} &
\multicolumn{1}{c}{$T_{\rm K}$} & \multicolumn{1}{c}{log(n$_{H_2}$)} &
\multicolumn{1}{c}{$N_{^{13}{\rm CO}}$ $^{b}$} &\multicolumn{1}{c}{\Nh2}\\
\multicolumn{1}{l}{} &
\multicolumn{1}{c}{K} & \multicolumn{1}{c}{log(\cmmt)} &
\multicolumn{1}{c}{10$^{16}$\cmmd} &\multicolumn{1}{c}{10$^{21}$\cmmd}\\
\hline
M+3.06+0.34 & 100&3-3.8 & 5.8-8  & 12-16\\
            & 20 & $>$3.3 & $\lsim$8 & $\lsim$16\\
M+1.56$-$0.30 & 100& 3-4$^{a}$ &2.8-10.2  &5.7-20 \\
            & 20 &3.5-4.5$^{a}$  & 2.8-8.1 & 5.6-16.2\\ 
\hline
\end{tabular}
     \begin{list}{}{}
        \item[$^{a}$] When \c18o is not detected, we have used,
            following      H\"uttemeister et al. (1998),  
           $n_{\rm H_2}$ $\sim$ 10$^{3-4}$ \cmmt when $T_{\rm K}$=100 K and,
           $n_{\rm H_2}$ $\sim$ 10$^{3.5-4.5}$ \cmmt when $T_{\rm K}$=20 K.
        \item[$^{b}$] Contains     the contribution from both velocity
               components.  Dispersion in the column densities 
                are due to  errors from the Gaussians
                fits to the spectra.
      \end{list}
\end{table}
\begin{table}[p]
\caption{Parameters of the best fits to the LWS spectra with
two gray bodies assuming $\alpha$=1 and $\Omega =\Omega_{\rm LWS}$:
 Temperatures, ratio of the opacity in the warmer component to the
 total opacity, and total visual extinction. Numbers in parentheses
are 1$\sigma$ errors        of the last significant digit.}
\begin{tabular}{lrrrr}
\hline
\multicolumn{1}{l}{Source} &
\multicolumn{1}{c}{$T_1$} &\multicolumn{1}{c}{$T_2$} 
&\multicolumn{1}{c}{$f$}& \multicolumn{1}{c}{$A_{\rm v}$}\\
\hline
M+3.06+0.34 &14(4) &24(2) &0.2(2)  & 30(20) \\
M+1.56$-$0.30 &15(4) &27(3) & 0.1(1)& 40(20)\\ 
\hline
\end{tabular}
\end{table}

\begin{table*}[p]
\caption{Para, ortho, and ortho-para rotational temperatures, 
OTPR and \h2 column densities. Numbers in parentheses are 
1$\sigma$ errors.} 
\begin{tabular}{lrrrrr}
\hline  
\multicolumn{1}{l}{Source} &
\multicolumn{1}{c}{$T_{\rm p}$} &\multicolumn{1}{c}{$T_{\rm o}$}
&\multicolumn{1}{c}{$T_{\rm op}$}& \multicolumn{1}{c}{OTPR}&
\multicolumn{1}{c}{\Nh2}\\
\multicolumn{1}{l}{} &
\multicolumn{1}{c}{(K)} &\multicolumn{1}{c}{(K)}
&\multicolumn{1}{c}{(K)}& \multicolumn{1}{c}{}&
\multicolumn{1}{c}{(10$^{21}$\cmmd)}\\
\hline
M+3.06+0.34  &260(30) &280(20) & 160(20) & 0.9(0.4) & 2.6(1.0)\\
M+1.56$-$0.30  &250(20) &270(20) & 160(20) & 1.0(0.4) & 2.1(0.8)\\
\hline
\end{tabular}
\end{table*}

\end{document}